\documentstyle[12pt]{article}
\begin{document}
\pagestyle{empty}
\baselineskip 18pt plus .1pt minus .1pt 
\def\r{\rightarrow}
\def\be{\begin{equation}}
\def\ee{\end{equation}}
\def\ben{\begin{eqnarray}}
\def\een{\end{eqnarray}}
\pagestyle{empty}
\begin{flushright}
US-00-02\\
April 19, 2000
\end{flushright}
\noindent
\begin{center}
{\large\bf{A Two Parameter Texture of Nearly Bi-maximal Neutrino  
Mixing }} 
\end{center}
\vskip .25in
\begin{center}
Ambar  Ghosal
\end{center}
\vskip .1in
\begin{center}
Department of Physics,
University of Shizuoka\\
52-1 Yada, Shizuoka-shi, Shizuoka 422 8526, Japan\\
\end{center}
\vskip 24pt
\noindent
We propose a texture of three generation 
Majorana-type neutrino mass 
matrix in terms of only two parameters which gives rise 
to nearly bi-maximal mixing 
angles. We also demonstrate an explicit realization 
of such type of neutrino mass-matrix in the context 
of an $SU(2)_L$$\times$$U(1)_Y$ model due to 
higher dimensional mass terms through the inclusion of
discrete $Z_3$$\times$$Z_4$ symmetry and two extra singlet 
Higgs fields.
\vskip 1.0in 
\noindent
PACS No. 12.60 Fr., 
14.60 Pq., 13.40 Em.\hfill
\begin{verbatim}
E-mail: ambar@snis.u-shizuoka-ken.ac.jp
        gp1195@mail.a.u-shizuoka-ken.ac.jp
\end{verbatim}
\begin{flushleft}
( To appear in Physical Review D )
\end{flushleft}
\newpage
\pagestyle{plain}
\setcounter{page}{2}
\noindent
{\bf{I.\,\, Introduction}}
\vskip .1in
\noindent
Evidence in favour of neutrino oscillation 
(as well as neutrino mass) has been 
provided by the Super-Kamiokande 
(SK) atmospheric 
neutrino experiment [1] through the measurement 
of magnitude and angular distribution of 
the $\nu_\mu$ flux produced in the atmosphere due to 
cosmic ray interactions. Observed depletion of 
$\nu_\mu$ flux in earth has been interpreted as the 
oscillation of $\nu_\mu$ to some other 
species of neutrino. In a two flavour neutrino 
oscillation scenario, oscillation between 
$\nu_\mu$ - $\nu_\tau$, the experimental result 
leds to maximal mixing between two species
$\rm{Sin^22\theta\ge 0.82}$ with a mass-squared 
difference 
$\Delta$$m^2_{atm}$$\sim$ 
$(5\times {10}^{-4}-6\times {10}^{-3})$ 
$\rm{eV^2}$.
Furthermore, recent result of SuperKamiokande experiment 
disfavours any large mixing between purely $\nu_\mu$ and 
$\nu_s$(sterile neutrino) at 99 $\%$ c.l.[2]. 
The solar neutrino experimental results [3] 
are also in concordance with the interpretation of 
atmospheric neutrino experimental result and the data 
provide the following values as 
$\Delta$$m^2_{e\mu}$$\sim$ 
$(0.8 - 2)\times {10}^{-5}\rm{eV^2}$, 
$\rm{Sin^22\theta}$ 
$\sim$ 1 (Large angle MSW solution) or 
$\Delta$$m^2_{e\mu}$$\sim$ 
$(0.5 - 6)\times {10}^{-10}\rm{eV^2}$, 
$\rm{Sin^22\theta}$ 
$\sim$ 1 (vacuum oscillation solution). 
Furthermore, the CHOOZ experimental result [4] 
gives the value of $\Delta m^2_{eX}< {10}^{-3}$
$\rm{eV^2}$ or $\rm{Sin^22\theta}_{eX}< 0.2$. In order 
to reconcile with the solar and atmospheric neutrino 
experimental results, a possible explanation known as 
 bi-maximal neutrino mixing is advocated [5], 
in which $\theta_{12}$=$\theta_{23}= 45^o$, and if, the 
CHOOZ experimental result is interpreted in terms of 
$\nu_e - \nu_\tau$ oscillation, 
then $\theta_{31}< 13^o$. Another scenario 
could still be possible if the solar neutrino 
experimental result is explained in terms of 
small angle MSW solution however, we have not address 
this scenario in the present work. 
In the prsent work, 
we propose a texture of Majorana-type neutrino mass matrix 
in terms of only two parameters considering only 
three generations of neutrinos. 
Two parameter texture of neutrino mass matrix has also been 
discussed earlier [6, 7, 8]. In Ref.6, with three light neutrinos, different zeroth order textures of both neutrino and charged lepton mass matrices 
has been proposed in view of the solar and atmospheric neutrino experimental results advocating the implication of flavor symmetry. A detailed analysis is found in Ref.7 where the implication of $L_e - L_\mu - L_\tau$ symmetry has been discussed to realize light neutrino mass both via see-saw mechanism and low energy effective theory. An investigation in this path has also been done in Ref.8 through the introduction of a partially conserved chiral $U(1)_{f_1}$$\times$ $U(1)_{f_2}$ symmetry with Standard model gauge group to generate both quark and lepton mass matrices. Apart from the successful description of quark and lepton mass matrices , however, in this model a large value of Higgs coupling of the term of dimension greater than four is needed to avoid the conflict between the minimization condition of the Higgs potential and the choice of low value of the VEV of an $SU(2)_L$ triplet Higgs field when vacuum oscillation solution of solar neutrino problem is considered in addition with the atmospheric neutrino experimental result. This problem is avoided in the present model by discarding any hard (dim $\geq$ 4) discrete symmetry violating term in the scalar potential.
\vskip .1in
\noindent
In this work,
we propose an explicit pattern of two parameter texture of neutrino mass matrix which 
gives 
rise to nearly bi-maximal neutrino mixing and also can accommodate 
the required mass-squared differences to explain the 
solar (by large angle MSW solution or by Vacuum oscillation 
) and atmospheric neutrino experimental results. 
Next, we demonstrate an explicit realization of the 
proposed texture within the framework of an $SU(2)_L$$\times$ 
$U(1)_Y$ model with an extended Higgs sector and discrete 
symmetry. The plan of the paper is as follows: Section IIcontains 
the proposed neutrino mass-matrix and its phenomenology. A model
accomplishes the proposed mass matrix is presented in Section III. 
Section IV contains summary of the present work.
\vskip 24pt
\noindent
{\bf{II. \,\, Neutrino\,\, Mass\,\, Matrix}}
\vskip .1in
\noindent
Before going into the details , first of all, we 
consider the charged lepton mass matrix is diagonal 
in flavor space. Consider now the following Majorana-type 
neutrino mass matrix 
with the basis of the leptonic fields 
($l_{1L}$, $l_{2L}$, $l_{3L}$) 
(where $l_{iL}$ have (2,1) quantum numbers under 
$SU(2)_L$$\times$$U(1)_Y$ gauge group, $i$ is the generation 
index)
\be
M_\nu = \pmatrix{0&a&a\cr
                 a&0&b\cr
                 a&b&0}
\ee
where $a$ and $b$ are two real model independent parameters and 
we consider $a\neq b$ so that $M_\nu$ contains at least two 
parameters.
Also it is to be noted that the absence of 
$\nu_e\nu_e$ mass term in the above neutrino 
mass matrix evades the bound on the Majorana-type neutrino 
due to $\beta\beta_{0\nu}$ decay. Moreover, the above texture 
admits no observable CP violating effect in the leptonic sector 
as the number of parameters is only two. The phases of $a$ and $b$ 
could easily be rotated away by redefining the leptonic fields.
The elements of $M_\nu$ can be generated either by 
radiative mechanism or by non-renormalizable mass operators.
We have not addressed here the see-saw type mass generation 
because in that case a judicious choice of Dirac-type 
neutrino mass matrix is necessary. 
Diagonalizing the neutrino 
mass matrix $M_\nu$ by an 
orthogonal transformation 
as $O^T M_\nu O$ = $M_D$ = 
Diag($-m_{\nu_1}$,$m_{\nu_2}$,$m_{\nu_3}$)  
where
\be
O = \pmatrix{c_{31}c_{12} & c_{31}s_{12} & s_{31}\cr
             -c_{23}s_{12} - s_{23}s_{31}c_{12} &  
             c_{12}c_{23} - s_{23}s_{31}
             s_{12} & s_{23}c_{31}\cr
             s_{23}s_{12} - c_{23}s_{31}c_{12} &  
             -s_{23}c_{12} - s_{31}s_{12}
             c_{23} & c_{23}c_{31}},
\ee
\noindent
we obtain the following values 
of the mixing angles as
\be 
\theta_{23}^\nu 
= -\frac{\pi}{4},\,\,  
\theta_{31}^\nu = 0,\,\, 
{\rm{tan^2}}\theta_{12}^\nu = 
\frac{m_{\nu_1}}{m_{\nu_2}} 
\ee
and the 
eigenvalues of the above mass matrix comes out as
\be
-m_{\nu_1} = \frac{b - x}{2} 
$$
$$
m_{\nu_2} = \frac{b + x}{2} 
$$
$$
-m_{\nu_3} = b 
\ee
where $x$ = $\sqrt{b^2 + 8 a^2}$.  
The sign of $m_{\nu_1}$ and $m_{\nu_2}$
 can be made positive by redefining lepton doublet 
fields. 
Furthermore, in terms of the three eigenvalues $m_{\nu_1}$, 
$m_{\nu_2}$ and  
$m_{\nu_3}$, the mixing matrix $O$ can be written as
\be
O =\pmatrix{c_{12} & s_{12} & 0\cr
             -\frac{1}{\sqrt 2}s_{12}&  
             \frac{1}{\sqrt 2}c_{12} & -\frac{1}{\sqrt 2}\cr
             -\frac{1}{\sqrt 2}s_{12}&  
             \frac{1}{\sqrt 2}c_{12}& \frac{1}{\sqrt 2}}
  =\pmatrix{{\sqrt{\frac{m_{\nu_2}}{m_{\nu_1} + m_{\nu_2}}}}
             & {\sqrt{\frac{m_{\nu_1}}{m_{\nu_1} + m_{\nu_2}}}}&
               0\cr
            -\frac{1}{\sqrt 2}{\sqrt{\frac{m_{\nu_1}}{m_{\nu_1} 
            + m_{\nu_2}}}}&
            \frac{1}{\sqrt 2}{\sqrt{\frac{m_{\nu_2}}{m_{\nu_1} 
            + m_{\nu_2}}}}& -\frac{1}{\sqrt 2}\cr
              -\frac{1}{\sqrt 2}{\sqrt{\frac{m_{\nu_1}}{m_{\nu_1} 
             + m_{\nu_2}}}}& \frac{1}{\sqrt 2}{\sqrt{\frac{m_{\nu_2}}
              {m_{\nu_1} 
               + m_{\nu_2}}}} & \frac{1}{\sqrt 2}}
\ee
In the limit 
$b$$\r$ 0 ,$\theta_{12}^\nu$$\r$$\frac{\pi}{4}$,
the two eigenvalues $m_{\nu_1}$ and $m_{\nu_2}$ 
become degenerate and 
we can achieve the exact bi-maximal 
neutrino mixing. In this situation, 
although we obtain the exact bi-maximal neutrino 
mixing however, the obtained eigenvalues 
$m_{\nu_1}$ = $m_{\nu_2}$  and 
$m_{\nu_3}$ = 0, can be fitted with 
either the solar or the 
atmospheric neutrino experimental result. 
Removal of degeneracy between the two 
eigenvlues require further higher order  
corrections. 
For our analysis, we set the value  
of $\Delta m^2_{21}$ = $\Delta m^2_{sol}$ which 
in turn sets the value of $\theta_{12}^\nu$. 
The value of $x$ 
depends on the hierarchical relation between 
$a$ and $b$ parameters 
which is manifested from the values 
of 
\be
\Delta m^2_{21} = bx 
\ee
and
\be
\Delta m^2_{23} = {\frac{1}{4}}(3b + x)(x - b). 
\ee
Now, if, $b^2 \gg 8a^2$, then 
the value of $x$ comes out 
as $x$  
$\simeq b$ and   
$\Delta m^2_{23}$$\simeq 0$,
$\Delta m^2_{21}$$\simeq b^2$,
 hence, in this 
case it is not possible 
to accommodate both the results of solar 
and atmospheric 
neutrino experiments. Thus, 
for a phenomenologically viable 
model, we have to consider 
the hierarchy $8a^2 \gg b^2$ and in this case 
$m_{\nu_1}$ is also become positive. 
The pattern of neutrino mass is presented in Figure I.
In this situation, we obtain, 
$\Delta m^2_{21}$ $\simeq$ $2{\sqrt 2} ab$, 
$\Delta m^2_{23}  
 $$\simeq$ 
$2a^2$. For a typical value of 
$\Delta m_{23}^2$$\simeq$ 
$4\times {10}^{-3}$ $\rm{eV^2}$  
which can explain the atmospheric 
neutrino deficits, 
we obtain 
$2a^2\simeq 4\times {10}^{-3}$ $\rm{eV^2}$. 
For a typical value of 
$\Delta m^2_{21}$$\simeq$$4\times {10}^{-10}$
$\rm{eV^2}$ which can explain 
the solar neutrino deficits due to  
vacuum oscillation, the value of $b^2$ 
comes out as $b^2\sim  
{10}^{-17}$$\rm{eV^2}$ 
whereas for the large angle MSW solution a typical 
value of 
$\Delta m^2_{21}$$\sim$${10}^{-5}$
$\rm{eV^2}$ the value of 
$b^2$ comes  out of the order of 
${10}^{-9}$ 
$\rm{eV^2}$.
The mixing angle $\theta_{12}^\nu$ comes out as 
$\tan^2\theta_{12}^\nu$ $\simeq$ $\frac{2a{\sqrt 2} - b}
{2a{\sqrt 2} + b}$ and since 
$a \gg b$, $\theta_{12}^\nu\r$ $45^o$, and, hence,  
there is no conflict to satisfy the value of 
$\theta_{12}^\nu$ well within the allowed 
range of the experimental value. 
\vskip .1in
\noindent
{\bf{III.\,\, A\,\, Model}}
\vskip .1in
\noindent
In this section, 
we demonstrate an explicit realization of the above 
neutrino mass matrix as well as a flavor 
diagonal charged lepton mass matrix within the framework 
of an 
$\rm{SU(2)_L\times 
U(1)_Y}$ model with two singlet 
Higgs fields   
 and discrete 
$\rm{Z_3\times Z_4}$ symmetry. 
The charged masses are arising in a similar way to 
Standard Model (SM) whereas neutrino masses are generated 
through non-renormalizable operators.
We have also discussed the situation when 
the mixing is exactly bi-maximal.
Instead of three almost 
degenerate 
neutrinos [9, 10], we obtain a 
hierarchical pattern of neutrino 
masses. 
To obtain a realistic low energy 
phenomenological model, several attempts 
have been made through the inclusion of discrete 
symmetry [11]. Recently, it has been shown [12] that non-
abelian discrete groups (such as dihedral groups ${\it D}_n$ 
, dicyclic groups ${\it Q}_{2n}$ )plays an attractive role 
to obtain required mixing pattern in the fermionic sector.
A recent work in this path has been done [13] through the 
inclusion of $U(1)$$\times$$Z_2$ symmetry in the flavor space 
to explain both the quark and leptonic sector mixing angles.
Although the question of embedding such symmetries under a large 
symmetry is still open , nevertheless, to understand from the 
low energy point of view, inclusion of discrete symmetry 
and extra matter fields is an attractive way.  
The discrete 
$\rm{Z_3\times Z_4}$ 
symmetry 
prohibits unwanted mass terms 
in the charged lepton 
and neutrino mass matrices in the present model. 
We consider soft 
discrete symmetry breaking terms in 
the scalar potential, 
which are also responsible to obtain 
non-zero values of the 
VEV's of the Higgs fields upon 
minimization of the scalar 
potential.
It is to be noted that in order to avoid conflict between the choice of VEV's of the Higgs fields ($\rho$ and $\xi$) with the minimization condition of the Higgs potential , we discard any hard discrete symmetry breaking term in the Higgs potential. Discrete symmetry invariant soft or hard terms will not cause hierarchical problem  as addressed in Ref.8.
 The Majorana neutrino 
masses are obtained 
due to explicit breaking of 
lepton number through higher 
dimensional terms. 
\begin{table}
\begin{center}
\begin{tabular}{c|c|c|c}
\hline
Fields & 
$SU(2)_L$ $\times$ $U(1)_Y$ & $Z_3$ & $Z_4$\\
\hline
{\underline{\rm{leptons}}}&&&\\
$l_{1L}$&(2, -1)& 1 & 1\\
$l_{2L}$&(2, -1)& $\omega$ & $-i$\\
$l_{3L}$&(2, -1)& $\omega$ & $i$\\
$e_R$&(1, -2)& $\omega^\star$ & 1\\
$\mu_{R}$&(1, -2)&1&-i\\
$\tau_{R}$&(1, -2)& 1& i\\
{\underline{\rm{Higgs}}}&&&\\
{\it{h}}&(2,1)&$\omega$&1\\
$\rho$&(1,0)&1&i\\
$\xi$&(1,0)&$\omega$&-1\\
\hline
\end{tabular}
\caption{Representation content of 
the lepton and Higgs fields considered 
in the present model. The 
generators of $Z_3$ 
and $Z_4$ groups are $\omega$ and $i$, respectively.}
\end{center}
\end{table}
The representation content of the leptonic fields 
and Higgs fields 
considered in the model is given in Table I. 
Apart from the standard model doublet $h$ 
Higgs field, we introduced another two
singlet Higgs $\xi$ and $\rho$ fields  to obtain two 
independent parameters for the neutrino 
sector.
\vskip .1in 
\noindent
The most general lepton-Higgs 
Yukawa interaction in the 
present model generating 
Majorana neutrino masses is given by 
\be
L_Y^\nu =  
{\frac{(l_{1L} l_{2L})hh \rho}{M_{f^2}}}
+ 
{\frac{(l_{1L} l_{3L})hh \rho^\star}{M_{f^2}}}
+
{\frac{(l_{2L} l_{3L})hh\xi^2}{M_{f^3}}}
\ee
and the Yukawa interaction which 
is responsible for generation 
of charged lepton masses is given by  
\be
L_Y^E =
 f_1 {\bar{l_{1L}}}e_R h
+
f_2 {\bar{l_{2L}}}\mu_R 
h
+
f_3{\bar{l_{3L}}} \tau_R h
+
H.c..
\ee
We  
consider $\rho$ is a complex scalar field whereas 
$\xi$ is a real scalar 
field.
The present model contains a large mass scale $M_f$, 
and for our analysis we set 
$M_f\sim M_{GUT}$ which is the highest scale 
considered in the present model. 
The  
VEV's, $\langle \xi \rangle$ 
and $\langle \rho \rangle$ are constrained by the solar 
and atmospheric neutrino 
experimental results.\hfill  
\vskip 24pt
\noindent
In order to avoid any zero values of the 
VEV's of the 
Higgs fields upon minimization of the 
scalar potential, we have 
to consider discrete symmetry 
breaking terms. Without 
going into the details of the scalar 
potential, this feature 
can be realized in the following way. 
In general, the scalar 
potential can be written 
as (keeping upto dim=4 terms)
\be
\rm{
V = Ay^4 + By^3 + Cy^2 + Dy + E
}
\ee
where '$y$' is the VEV of 
any Higgs field and A, B, C, D, E are 
generic couplings of the terms 
contained in the scalar potential. 
Minimizing the scalar  
potential w.r.t. '$y$', we obtain
\be
\rm{
V^\prime = A^\prime y^3 + 
B^\prime y^2 + C^\prime y + D
}
\ee
Eqn.(10) reflects the fact 
that as long as $\rm{D}\neq 0$, and 
$A^\prime$ or $B^\prime$ or $C^\prime$ 
is not equal to zero, we 
will get non-zero solutions for '$y$'. 
Thus, in order to obtain $y\neq 0$ 
solution, it is necessary 
to retain the terms 
with generic 
coefficients D and  
$A^\prime$ or $B^\prime$ or $C^\prime$. 
In the present model, 
both the discrete symmetry breaking terms 
soft and hard, correspond to the term with 
coefficient D. Discarding hard symmetry breaking 
terms, we retain soft discrete symmetry breaking 
terms, and, hence, none of the VEV is zero upon 
minimization of the scalar potential.\hfill   
\vskip .1in    
\noindent
Let us look at the leptonic 
sector of the prsent model. Substituting the 
VEV's of the Higgs fields appeared in Eqn.(9), we 
obtain flavor diagonal charged lepton mass matrix as
\be
M_E = \pmatrix{d&0&0\cr
               0&e&0\cr
               0&0&f}
\ee
where $d = f_1 \langle h \rangle $,  
$e = f_2 \langle h \rangle$ and   
$f = f_3 \langle h \rangle$ and substituting the 
VEV's of $\xi$, 
$h$ and $\rho$ Higgs fields in Eqn.(8), 
we get the Majorana-type  
neutrino mass matrix as follows: 
\be
M_\nu = \pmatrix{0&a&a\cr
               a&0&b\cr
               a&b&0}
\ee
where\footnote{The discrete symmetry invariant 
$\nu_e\nu_e$ mass term appears in the 
present model at $M_f^5$ order 
which is naturally vanishingly small.} 
$a = \frac{\langle\rm{h}\rangle^2 
\langle\rho\rangle}{M_f^2}$,  
$b = \frac{<\xi>^2<\rm{h}>^2}{M_f^3}$. 
The parameter $a$ can fitted with the value 
$\Delta m^2_{23}$ $\simeq$ 
$2 a^2$ $\simeq$ $4\times {10}^{-3}$ $\rm{e V^2}$ 
which explains atmospheric 
neutrino experimental data by setting 
$M_f$$\sim$ ${10}^{12}$ GeV, 
$\langle h \rangle$$\simeq$ 174 GeV 
and $\langle\rho\rangle$ 
$\simeq$${10}^{11}$ GeV. Using the same 
values of $M_f$ and $\langle h \rangle$, 
it is possible to set 
the value of $b$ as $b^2$$\simeq$
${10}^{-17}$$\rm{eV^2}$ through 
the choice of $\langle\xi\rangle$
$\simeq$${10}^7$ GeV in order to explain 
the solar neutrino 
experimental results due to vacuum oscillation 
solution. For both the cases , the mixing angle 
$\theta_{12}^\nu$ ( given in Eqn.(3)) 
comes out as nearly maximal.
For large angle MSW solution, a typical value 
$\Delta m^2_{21}$$\sim$ ${10}^{-5}$ $\rm{e V^2}$
gives rise to 
$b^2$$\simeq$ ${10}^{-9}$$\rm{eV^2}$ for 
$\langle\xi\rangle\simeq 2\times {10}^9$ GeV.\hfill
\vskip .1in
\noindent
{\bf{IV. \,\, Summary}}
\vskip .1in
\noindent
In summary, 
we propose a texture of Majorana-type neutrino 
mass-matrix which gives rise to nearly bi-maximal 
neutrino mixing in a natural way as well as required 
mass-squared differences in order to explain the 
solar and atmospheric neutrino experimental results. The 
elements of the mass-matrix could be generated either 
by radiative mechanism or by the use of non-renormalizable 
operators and, thus, those elements are model independent. 
The proposed neutrino mass-matrix gives rise to the eigenvalues 
of the three neutrino masses as 
$m_{\nu_1}$ $\simeq$ $m_{\nu_2}$ $\gg$ $m_{\nu_3}$ 
which ends up to an hierarchy between three neutrino 
mass-squared differences as $m_{23}^2$ $\gg$ $m_{21}^2$.
We demonstrate an explicit realization of the proposed 
mass-matrix due to non-renormalizable mass operators 
in the context of an $SU(2)_L$$\times$$U(1)_Y$ model through 
the inclusion of two extra singlet Higgs fields and 
discrete $Z_3$$\times$$Z_4$ symmetry. With a suitable 
choice model parameters the required mass-squared differences can 
be accommodated in order to expalin the solar 
(both large angle MSW solution 
and Vacuum oscillation) and atmospheric neutrino 
experimental results.
\vskip .25in   
Author acknowledges Yoshio Koide 
for many helpful 
suggestions,discussions and careful reading of the 
manuscript and also thankful to Masaki Yoshimura 
for helpful comments. Part of this work has been done 
under the financial support of The Japan Society For 
The Promotion 
of Science, Govt. of Japan.
\newpage
\begin{center}
{\large\bf{References}}
\end{center}
\begin{enumerate}
\item Super - Kamiokande Collaboration, 
 Y. Fukuda et al., Phys. Lett. B433, 
(1998) 9, {\it{ibid}} 436, (1998) 33, 
Kamiokande  Collaboration, 
S. Hatakeyama et al., Phys. Rev. Lett. 
81 
 (1998) 2016, T. Kajita, Talk presentd at 
 'Neutrino 98', Takayama, Japan, (1998). 

\item Y. Obayashi,  `Neutrino Oscillations and their 
 origin` , Fujiyoshida, Japan, Feb 11 - 13 (2000), University 
Academy Press, Tokyo.

\item R. Davis, Prog. Part. Nucl. Phys. 32 (1994) 13; 
Y. Fukuda et al., Phys. Rev. Lett. 77 (1996) 1683, 
P. Anselmann et al., Phys. Lett. B357, (1995) 237, 
{\it{ibid}} B361, (1996) 235.
 
\item M. Appollonio et al. Phys. Lett. B420 (1998) 397.

\item V. Barger, S. Pakvasa, 
T. J. Weiller and  K. Whisnant, 
hep-ph/9806387, 
B. C. Allanach, hep-ph/9806294,
D. V. Ahluwalia, Mod. Phys. Lett. A 13 (1998) 2249, 
I. Stancu and D. V. Ahluwalia, Phys. Lett. B460, 431 (1999),
 V. Barger, T. J. Weiller and K. Whisnant, 
hep-ph/9807319, J. Elwood, N. Irges and 
P. Ramond, Phys. Rev. Lett. 81, (1998) 5064, 
hep-ph/9807228, E. Ma, 
Phys. Lett. B442, (1998) 238,hep-ph/
9807386, hep-ph/9902392, 
G. Alterelli and F. Feruglio, 
Phys. Lett. B439 (1998) 112,hep-ph/9807353, 
Y. Nomura and T. Yanagida, Phys. Rev. 
D59 (1999) 017303, 
 hep-ph/
9807325,  A. Joshipura, hep-ph /98 08261, 
A. S. Joshipura and S. Rindani, 
hep-ph/9811252, 
K. Oda et al., Phys. Rev. D59, (1999) 055001, 
hep-ph/9808241, H. Fritzsch and 
Z. Xing, hep-ph/9808272, J. Ellis 
et al., hep-ph/
9808301, A. Joshipura and S. Vempati, hep-ph/
9808232, U. Sarkar, Phys. Rev. 
D59, (1999) 037302,
hep-ph/9808277,   
H. Georgi and S. Glashow , hep-ph/ 9808293;
A. Baltz, A. S. Goldhaber and M. Goldhaber, 
Phys. Rev. Lett. 81, (1998) 5730, 
hep-ph/ 9806540, 
M. Jezabek and A. Sumino, 
Phys. Lett. B440, (1998) 327, hep-ph/ 9807310,  
S. Davidson and S. F. King, 
Phys. Lett. B445 (1998) 191,
hep-ph/ 9808296,
 K. Kang, S. K. Kang, C. S. Kim and  
 S. M. Kim, 
 hep-ph/ 9808419, S. Mohanty, 
D. P. Roy and  U. Sarkar, 
Phys. Lett. B445, (1998) 185,
hep-ph/9808451;   
E. Ma, U. Sarkar and D. P. Roy, Phys. Lett. 
B444, (1998) 391, hep-ph/9810309,
B. Brahmachari, hep-ph/9808331,
R. N. Mohapatra and S. Nussinov, 
Phys. Lett B441, (1998) 299, 
hep-ph/ 9808301, Phys. Rev. 
D60, (1999) 013002, hep-ph/ 9809415,  
A. Ghosal, hep-ph/9903497, hep-ph/9905470, 
R. Barbieri, L.J.Hall and A. Strumia, 
Phys. Lett. B445 (1999) 407, 
hep-ph/9808333, Y. Grossmann, Y. Nir and 
Y. Shadmi, JHEP, 9810 (1998) 007, 
hep-ph/9808355, 
C. Jarlskog, M. Matsuda, S. Skadhauge and M. 
Tanimoto, 
Phys. Lett. B449, (1999) 240, 
hep-ph/9812282, S. M. Bilenky and 
C. Giunti, hep-ph/9802201, 
C. Giunti, Phys. Rev. D59, (1999) 077301,  
hep-ph/ 
9810272, 
M. Fukugita, M. Tanimoto and T. Yanagida, 
Phys. Rev. D57 (1998) 4429, 
S. K. Kang and C. S. Kim, 
Phys. Rev. D59, (1999) 091302,  
R. N. Mohapatra, A. Perez-Lorenzana and C.A. de S. Pires, 
Phys. Lett. B474 (2000) 355, A. Aranda, 
C. D. Carone and R. F. Lebed , Phys. Lett. B474 
(2000) 170, hep-ph/0002044,
H. B. Benaoum and S. Nasri, hep-ph/9906232, 
C. H. Albright and S. M. Barr, hep-ph/9906297. 

\item R. Barbieri, L. J. Hall and A. Strumia, Phys. Lett. B445, 407 (1999), hep-ph/9808333.

\item R. Barbieri et al., JHEP 9812 : 017, (1998), hep-ph/9807235.

\item C. D. Froggatt, M. Gibson and H. B. Nielsen, hep-ph/9811265,
Phys. Lett. B446, 256 (1999). 

\item D. O. Caldwell and R. N. Mohapatra, 
Phys. Rev. D50, (1994) 3477, A.S.Joshipura, 
Z.Phys. C 64 (1994) 31,
D. G. Lee and R. N. Mohapatra, Phys. Lett. B329, 
(1994) 463, P.Bamert and C.P.Burgess, Phys. Lett. B329, 
(1994) 289; E. Ma, hep-ph/9812344, 
Y. L. Wu, hep-ph/9810491, 
hep-ph/9901245, hep-ph/9901320, 
hep-ph/9905222. 

\item A. S. Joshipura, Phys. Rev. D51(1995), 1321, 
A. Ioannissyan and J. W. F. Valle, Phys. Lett.  
B332, (1994) 93, A. Ghosal, Phys. Lett. B398, (1997) 315, 
A. K. Ray and S. Sarkar, Phys. Rev. D58, (1998) 055010. 

\item K. Fukuura, et. al., hep-ph/9909415, T. Miura, E. Takasugi 
and M. Yoshimura, hep-ph/0003139, M. Tanimoto, hep-ph/0001306, 
Y. Koide, Phys. Rev. D60, 077301, 1999.

\item P. H. Frampton and A. Rasin, hep-ph/9910522. 

\item M. S. Berger and K. Siyeon, hep-ph/0003121. 
\end{enumerate}
\newpage
\begin{picture}(450,160)(0,0)
\setlength{\unitlength}{1pt}
{  
\thicklines
\put(100,100){\rule{80pt}{2pt}}
\put(100,80){\rule{80pt}{2pt}}
\put(100,30){\rule{80pt}{2pt}}
}
\put(50,95){  $\mbox{\boldmath$m_{\nu_2}$}$}
\put(50,75){  $\mbox{\boldmath$m_{\nu_1}$}$}
\put(50,25){  $\mbox{\boldmath$m_{\nu_3}$}$}
\put(150,55){\thicklines\vector(0,1){23}}
\put(150,55){\thicklines\vector(0,-1){23}}
\put(170,45){\bf Atm}
{ 
\put(185,90){\thicklines\vector(0,1){8}}
\put(185,90){\thicklines\vector(0,-1){8}}
\put(190,90){\bf {Solar}}
}
\end{picture}
\vskip 1in
\begin{center}
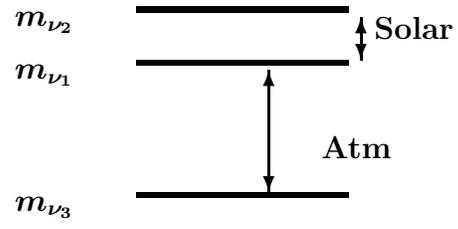
{\bf{FIG. I.}} Neutrino mass spectrum in the present model.
\end{center}
\end{document}